\documentclass[a4paper,12pt]{article}
\usepackage{citeref}

\input cohmacros.sty

\def\at#1{[*** \att #1 ***]}  
\def\at#1{} 

\advance\textheight2.7cm        
\advance\topmargin-2.0cm
\advance\textwidth2.6cm
\advance\evensidemargin-1.6cm
\advance\oddsidemargin-1.6cm

\def\B {{\bf B}}
\def\E {{\bf E}}

\begin{document}

\vspace*{-2cm}
\begin{center}
{\LARGE \bf Foundations of quantum physics} \\[4mm]

{\LARGE \bf  V. Coherent foundations} \\

\vspace{1cm}

\centerline{\sl {\large \bf Arnold Neumaier}}

\vspace{0.5cm}

\centerline{\sl Fakult\"at f\"ur Mathematik, Universit\"at Wien}
\centerline{\sl Oskar-Morgenstern-Platz 1, A-1090 Wien, Austria}
\centerline{\sl email: Arnold.Neumaier@univie.ac.at}
\centerline{\sl \url{http://www.mat.univie.ac.at/~neum}}

\end{center}


\hfill May 19, 2019

\vspace{0.5cm}

\bigskip
\bfi{Abstract.}
This paper is a programmatic article presenting an outline of a new 
view of the foundations of quantum mechanics and quantum field theory. 
In short, the proposed foundations are given by the following 
statements:

\pt
Coherent quantum physics is physics in terms of a coherent space 
consisting of a line bundle over a classical phase space and an 
appropriate coherent product. 

\pt
The kinematical structure of quantum physics and the meaning of the
fundamental quantum observables are given by the symmetries of this 
coherent space, their infinitesimal generators, and associated 
operators on the quantum space of the coherent space. 

\pt
The connection of quantum physics to experiment is given through the
thermal interpretation. The dynamics of quantum physics is given (for
isolated systems) by the Ehrenfest equations for q-expectations.

\vfill
For the discussion of questions related to this paper, please use
the discussion forum \\
\url{https://www.physicsoverflow.org}.

\newpage
\tableofcontents 

\vspace{2cm}

\section{Introduction}

This paper, the fifth of a series of papers 
\cite{Neu.Ifound,Neu.IIfound,Neu.IIIfound,Neu.IVfound} on the 
foundations of quantum physics, and the third one of a series of papers 
(\sca{Neumaier} \cite{cohSpaces.html}) on coherent spaces and their 
applications, presents a new view of the foundations for 
quantum mechanics and quantum field theory, highlighting the 
problems and proposing solutions. In short, the proposed 
\bfi{coherent foundations} are given by the following statements, made 
precise later:

Coherent quantum physics is physics in terms of a coherent space 
consisting of a line bundle over a classical phase space and an 
appropriate coherent product. The kinematical structure of quantum 
physics and the meaning of the quantum observables\footnote{
In the following, these will be called \bf{quantities} or q-observables 
to distinguish them from observables in the operational sense of 
numbers obtainable from observation. Similarly, we use at places 
q-expectation for the expectation value of quantities.
} 
are given by the symmetries of this coherent space, their infinitesimal 
generators, and associated operators on the quantum space of the 
coherent space. 

The connection of quantum physics to experiment is given through the
thermal interpretation. The dynamics of quantum physics is given (for
isolated systems) by the Ehrenfest equations for q-expectations.

The coherent foundations proposed here in a programmatic way resolve
the problems with the traditional presentation of quantum mechanics 
discussed in Part I \cite{Neu.Ifound}.

\bigskip

This paper is a programmatic overview article containing the main ideas 
on coherent spaces and their relation to quantum physics, not the 
precise concepts. These are defined and studied in depth in other 
papers of the series on coherent spaces, beginning with
\sca{Neumaier} \cite{Neu.cohPos} and \sca{Neumaier \& Ghaani Farashahi}
\cite{NeuF.cohQuant}. See also the exposition at the web site
\sca{Neumaier} \cite{cohSpaces.html}.

Section \ref{s.coh} gives rigorous definitions of the most basic
concepts and results on coherent spaces, without attempting to be 
comprehensive, and (together with the next section) a general outline 
of a coherent quantum physics, telling the main points of the story with
as few formulas and conceptual details as justifiable.

Section \ref{s.cohQuant} introduces the concept of symmetries 
(invertible coherent maps) of coherent spaces and associated 
quantization procedures. 
This leads to quantum dynamics, which in special (completely integrable)
situations can be solved in closed form in terms of classical motions 
on the underlying coherent space, if the latter has a compatible 
manifold structure. Spectral issues can in favorable cases be handled 
in terms of dynamical Lie algebras. Close relations to concepts from 
geometric quantization and K\"ahler manifolds are pointed out. 

In Section \ref{s.TILie}, we rephrase the formal 
essentials of the thermal interpretation in a slightly generalized more 
abstract setting, to emphasize the essential mathematical features and 
the close analogy between classical and quantum physics. We show how
the coherent variational principle (the Dirac--Frenkel procedure 
applied to coherent states) can be used to show that in coarse-grained 
approximations that only track a number of relevant variables, quantum 
mehcnaics exhibits chaotic behavior that, according to the thermal 
interpretation, is responsible for the probabilistic features of 
quantum mechanics. 

The final Section \ref{s.fieldTh} defines the meaning of the notion of 
a field in the abstract setting of Section \ref{s.TILie} and shows how 
coherent spaces may be used to define relativistic quantum field 
theories.

\bigskip

The puzzle of making sense of the foundations of quantum physics held my
attention for many years. Around 2003, I discovered that group 
coherent states are for many purposes very useful objects; before, they 
were for me just a facet that physicists studied who needed them for 
quantum optics. In 2007, I realized that apparently all of quantum 
mechanics and quantum field theory can be profitably cast into this 
form, and that coherent states may provide better theoretical 
foundations for quantum mechanics and quantum field theory than the
current Fock space approach. Since then I have been putting piece by 
piece into the new framework, and always found (after some work) 
everything nicely fitting.
With each new piece in place, I got insights about how to interpret
everything, and things got simpler and simpler as I proceeded. Or
rather, more and more complicated things became understandable without
significant increase of complexity in the new picture. Everything
became much more transparent and intuitive than the traditional
mental picture of quantum physics was.

In the bibliography, the number(s) after each reference give the page 
number(s) where it is cited.

\bigskip

{\bf Acknowledgments.}
Earlier versions of this paper benefitted from discussions with
Rahel Kn\"opfel and Mike Mowbray.

\section{Coherent spaces}\label{s.coh}

Coherent quantum physics is quantum physics in terms of a
coherent space consisting of a classical phase space and an appropriate
coherent product. The kinematical structure and the meaning of the
quantitys are given by the symmetries (invertible coherent maps) of
the coherent space.

This section gives rigorous definitions of the most basic
concepts and results, without attempting to be comprehensive,
and (together with the next section) a general outline of a
coherent quantum physics, telling the main points of the story with
as few formulas and conceptual details as justifiable. Unexplained
details can be found in my papers on coherent spaces (and the
references given there). Two of these papers (\sca{Neumaier}
\cite{Neu.cohPos} and \sca{Neumaier \& Ghaani Farashahi}
\cite{NeuF.cohQuant}) are already publicly available; others are in
preparation and will become available at my web site
\cite{cohSpaces.html}.

\bigskip

Coherent spaces are a novel mathematical concept, a nonlinear version
of Hilbert spaces. They combine the rich,
often highly characteristic variety of symmetries of traditional
geometric structures with the computational tractability of
traditional tools from numerical analysis and statistics.

To get the axioms of a coherent space from those of a Hilbert space,
the vector space axioms are dropped while the notion of inner product
and its properties is kept.
Every subset of a real or complex Hilbert space may be viewed as a
coherent space.
Symmetries induced by orthogonal resp. unitary transformations become
symmetries of the coherent space.

Conversely, every coherent space can be canonically embedded into a
complex Hilbert space (namely its quantum space) in such a way that all
its symmetries are realized by unitary transformations.
Thus, in a way, the theory of coherent spaces is just the theory of
subsets of a Hilbert space and their symmetries.
However, just as it pays to study the properties of manifolds
independently of their embedding into a Euclidean space, so it appears
fruitful to study the properties of coherent spaces independent of
their embedding into a Hilbert space.

There are close connections to reproducing kernel Hilbert
spaces, leading to numerous applications in quantum physics,
complex analysis, statistics, and stochastic processes.

One of the strengths of the coherent space approach is that it makes
many different things look alike, and stays close to actual
computations. There are so many applications in physics and elsewhere
that pointing them all out will take a whole book to write\ldots

Coherent states and squeezed states in quantum optics, mean field
calculations in statistical mechanics, Hartree--Fock calculations for
the electronic states of atoms, semiclassical limits, integrable
systems all belong here. As will be shown in later papers from this
series, most computational techniques in quantum physics can be
profitably phrased in terms of coherent spaces.

\subsection{Coherent spaces}

Fundamental is the notion of a coherent space. It is a nonlinear version
of the notion of a complex Hilbert space:
The vector space axioms are dropped while the notion of
inner product, now called a coherent product, is kept.
Every coherent space can be embedded into a Hilbert space
extending the coherent product to an inner product.

In informal, traditional terms, a coherent space is roughly a set $Z$
whose elements label certain vectors, called \bfi{coherent states} of a
Hilbert space.
The quantum space of $Z$ is the closed subspace formed by the
limits of linear combinations of coherent states.

However, one can characterize this situation independent of a Hilbert
space setting. Then a coherent space is a set $Z$ equipped
(among others) with a so-called coherent product that assigns to
any two points $z,z'\in Z$ a complex number $K(z,z')$ satisfying
certain coherence properties. The coherent product is essentially the
inner product in the quantum space of the coherent states with the
corresponding classical labels.

More formally, a \bfi{Euclidean space} is a complex vector space $\Hz$
with a binary operation that assigns to $\phi,\psi\in\Hz$ the
\bfi{Hermitian inner product} $\phi^*\psi\in\Cz$, antilinear in the
first and linear in the second argument, such that
\lbeq{e.her}
\ol{\phi^*\psi}=\psi^*\phi,
\eeq
\lbeq{e.def}
\psi^*\psi>0 \Forall \psi\in\Hz\setminus\{0\}.
\eeq
In physics, one usually writes $\<\phi|\psi\>$ in place of $\phi^*\psi$,
but we reserve this bra-ket notation exclusively for coherent states,
as defined below.
$\Hz$ has a natural locally convex topology in which the inner product
and any linear functional is continuous, and is naturally embedded into 
its antidual $\Hz^\*$, the space of antilinear functionals on $\Hz$. 
The Hilbert space completion $\ol\Hz$ sits between these two spaces,
\[
\Hz\subseteq\ol\Hz\subseteq\Hz^\*.
\]
$\Linx \Hz$ denotes the space of linear mappings from
$\Hz$ to $\Hz^\*$; they are automatically continuous.

A \bfi{coherent space} is a nonempty set $Z$ with a distinguished
function $K:Z\times Z\to\Cz$, called the \bfi{coherent product}, such
that
\lbeq{e.Kherm}
\ol{K(z,z')}=K(z',z),
\eeq
and for all $z_1,\ldots,z_n\in Z$, the $n\times n$ matrix $G$ with
entries $G_{jk}=K(z_j,z_k)$ is positive semidefinite.

The \bfi{distance} (\sca{Parthasarathy \& Schmidt} \cite{ParS}) 
\lbeq{e.dist}
d(z,z'):=\sqrt{K( z,z)+K( z',z')-2\re K( z,z')}
\eeq
of two points $z,z'\in Z$ is nonnegative and satisfies the triangle
inequality.
The distance is a metric precisely when the coherent space is
\bfi{nondegenerate}, i.e., iff
\[
K(z'',z')=K(z,z')~~\forall~ z'\in Z \implies z'' = z.
\]
In the resulting topology, the coherent product is continuous.

A \bfi{coherent manifold} is a smooth ($=C^\infty$) real manifold $Z$
with a smooth coherent product $K:Z\times Z\to \Cz$ with which $Z$ is a
coherent space.
In a nondegenerate coherent manifold, the infinitesimal distance equips
the manifold with a canonical Riemannian metric.

A \bfi{quantum space} $\Qz(Z)$ of $Z$ is a Euclidean space
spanned (algebraically) by a distinguished set of vectors $|z\>$
($z\in Z$) called \bfi{coherent states} satisfying
\lbeq{e.cohProd}
\<z|z'\> ~=~ K(z,z') \for z,z'\in Z
\eeq
with the linear functionals
\[
\<z|:=|z\>^*
\]
acting on $\Qz(Z)$. Coherent states with distinct labels are distinct
iff $Z$ is nondegenerate.

A construction of \sca{Aronszajn}  \cite{Aro0,Aro} (attributed by him to
\sca{Moore} \cite{Moo}), usually phrased in terms of reproducing kernel
Hilbert spaces, proves the following basic result.

{\bf Moore--Aronszajn Theorem.}
{\it Every coherent space has a quantum space. It is unique
up to isometry.} 

The antidual  $\Qz^\*(Z):=\Qz(Z)^\*$ of the quantum space $\Qz(Z)$
is called the \bfi{augmented quantum space}. It contains the
\bfi{completed quantum space} $\ol\Qz(Z)$, the Hilbert space
completion of $\Qz(Z)$,
\[
\Qz(Z)\subseteq\ol\Qz(Z)\subseteq\Qz^\*(Z).
\]
In quantum mechanical applications, $\ol\Qz(Z)$ is the Hilbert space
containing the pure states, while $\Qz^\*(Z)$ also contains
unnormalizable wave functions.

Constructing Hilbert spaces from a coherent space and its coherent
product is much more flexible, and hence more powerful, than the
standard approach of constructing Hilbert spaces from a function space
and a measure on it. Virtually
every Hilbert space arising in quantum mechanical practice can be neatly
constructed as the quantum space of an appropriate coherent space;
the preceding examples gave the first bits of evidence of this.

In a quantum mechanical context, $Z$ is a classical phase space or
extended phase space -- typically a symplectic manifold, a Poisson
manifold, or a circle or line bundle over such a manifold that
incorporates the classical action variable (encoding the Berry phase
under quantization). For example, the Aharonov--Bohm effect \cite{AhaB}
needs the bundle formulation. A canonical symplectic form is determined
by the coherent product.
The precise relationship is the subject of \bfi{geometric
quantization}, loosely outlined in Subsection \ref{ss.geomQuant}.

This provides a classical view of the system. On the other hand, the 
coherent product also determines its quantum space, whose completion 
$\ol\Qz(Z)$ is the Hilbert space of quantum mechanical state vectors. 
This provides a quantum view of the system.

Thus coherent spaces allow both a classical and a quantum view of
the same system. The two views are closely related, as the phase space
points $z\in Z$ label a family of \bfi{coherent states} $|z\>$, special
vectors in the quantum space for which the inner product takes the
simple form
\lbeq{e.cip}
\<z|z'\>=K(z,z').
\eeq
Thus in some sense, the classical phase space and the quantum Hilbert
space coexist in the framework of coherent spaces. The classical phase
space is a quotient space of $Z$ under the equivalence relation that
identifies points whose corresponding coherent states differ only by a
scale factor. Thus points in the phase space are in 1-1 correspondence
with equivalence classes of points of $Z$, hence equivalence classes of
labels of coherent states. The quantum space is the completion of the
space spanned by all coherent states. It is a Hilbert space that
can be realized as a space of functions on $Z$; the coherent states
$|z\>$ are essentially the functions that map $z'\in Z$ to the coherent
product $K(z,z')$.

If we regard $Z$ as a classical phase space, as often adequate, the
functions
\[
\wh \psi (z):=\psi^T|z\>,~~~\psi\in \Qz(Z).
\]
are those classical phase space functions that have an immediate 
quantum meaning.
Note that $\Qz^\*(Z)$ consists of all complex-valued maps on $Z$ that
are continuous in the natural weak topology induced by the coherent
product.

Glauber coherent states (mentioned before)
are a particular instantiation of this concept.
A more trivial case to keep in mind is to label {\em all} vectors in
finite-dimensional Hilbert space $\Cz^n$, so that $Z=\Cz^n$ and
$\<z|z'\>=K(\ol z,z')$ with
\lbeq{e.trivial}
K(z,z'):=z^*z'=\sum_k \ol z_kz'_k.
\eeq
This extends to infinite dimensions
(the usual case in most of quantum physics) by replacing the sum by
an appropriate integral, and shows that the traditional way of looking
at Hilbert spaces can be fully accommodated with such a coherent space.
However, this choice is poor from the point of view of the
classical-quantum correspondence. As we shall see, there are far better
choices, leading to a much increased flexibility compared to the
traditional approach of defining Hilbert spaces by giving the inner
product as a sum or integral. More importantly, as one works most of
the time in $Z$ and very little explicitly in the quantum space, one
can often use classical intuition in quantum situations, and the economy
of classical computations is often preserved.

Finite linear combinations of coherent states form a dense subspace
$\Qz(Z)$ of the Hilbert space $\ol\Qz(Z)$. This implies that all
quantum mechanical calculations, usually done in an orthonormal basis,
can also be done on the basis of coherent states, and often far more
efficiently.
Most conceptual issues can be discussed in coherent terms, too.
This makes the closeness to a classical description very plain,
and removes most of the mystery of quantum physics.

The simplest classical systems have a finite number $N$ of states,
corresponding to a phase space $Z$ with $N$ elements.
Their dynamics is that of a hopping process, a \bfi{continuous
time Markov chain} determined by consistently specifying transition
rates for hopping from one state to another. More complex
classical systems have phase spaces $Z$ that are finite-dimensional
manifolds when there are only finitely many degrees of freedom.
In particular, this is the arena of \bfi{classical mechanics} of point
particles, where $Z$ is a symplectic manifold, or more generally a
Poisson manifold. The deterministic dynamics is defined on $Z$ by
Hamilton's equations, equivalently on phase space functions by means of
the Poisson bracket.
Finally, in \bfi{classical field theory}, the phase space $Z$ is an
infinite-dimensional space of fields in 3-dimensional space, the
deterministic dynamics on $Z$ is described by partial differential
equations. Often an equivalent dynamics on phase space functions
(now functions on fields) is given in terms of an appropriate Poisson
structure on $Z$.

The simplest quantum systems have a finite number $N$ of levels,
corresponding to a Hilbert space of dimension $N$. We may consider them
as the quantum version of a Markov chain; this corresponds to picking
an orthonormal basis of $N$ pointer vectors $|z\>$ and declaring the
coherent product to be $K(z,z'):=\<z|z'\>=\delta_{zz'}$, thus creating a
coherent space $Z$ with $N$ elements. However, a 2-level quantum system
also models a \bfi{spinning electron} at rest in its ground state. Here
the appropriate classical analogue is not the counterintuitive two state
(up-down) model which depends on a distinguished direction and hence
sacrifices the spherical symmetry of the electron, but a 2-sphere in
$\Rz^3$, the phase space of a classical spinning top. To account for
the nonintegral spin of the electron, we should in fact take as
classical phase space the double cover of the 2-sphere, given by the
unit sphere
\[
Z=\{z\in\Cz^2\mid z^*z=1\}
\]
in $\Cz^2$. (The double cover is the so-called Hopf fibration, a
nontrivial object.)
The 3-sphere is the same thing as the unit sphere in $\Cz^2$ written
in real coordinates. The discussion of the Hopf fibration in terms of
quaternions can be interpreted in terms of Pauli matrices, giving
the traditional approach to 2-level systems. In terms of coherent
states, all these technicalities are hidden -- one has the quantum
space without having to bother about the latter. This economy of
coherent states becomes more pronounced in more complicated models,
which is the most important one of the reasons why they are studied
here.

To get the correct
2-state quantum space, we need to take the trivial coherent product
\gzit{e.trivial} restricted to $Z$. Remarkably, the case of a particle
of \bfi{higher spin} $j$ has the same phase space, with the coherent
product only slightly changed to
\lbeq{e.cohpsj}
K(z,z'):=(z^Tz')^{2j+1}.
\eeq
Equally remarkably, the coherence conditions are satisfied for this
coherent product only if $j=0,\half,1,\frac{3}{2},\dots$, thus
naturally accounting for the fact that spin is quantized.

In contrast, accounting for arbitrary spin in the traditional fashion
based on a $(2j+1)$-level system requires a significant amount of
machinery already to define the representation.

\subsection{Examples} \label{ss.IntroRems}

{\bf Example: Klauder spaces.}
The Klauder space $KL[V]$ over the Euclidean space $V$ is
the coherent manifold $Z=\Cz\times V$ of pairs
$z:=[z_0,\z]\in \Cz\times V$ with coherent product
$K(z,z'):=e^{\ol z_0 +z_0'+\z^*\z'}$.
($KL[\Cz]$ is essentially in \sca{Klauder} \cite{Kla}. Its coherent
states are precisely the nonzero multiples of those discovered by
\sca{Schr\"odinger} \cite{Schr}.)
As shown in detail in \sca{Neumaier \& Ghaani Farashahi}
\cite{NeuF.cohQuant}, where coherent construction of creation 
annihilation operators together with their properties are derived, 
the quantum spaces of Klauder spaces are essentially the
\bfi{Fock spaces} introduced by \sca{Fock} \cite{Foc} in the context of
\bfi{quantum field theory}. They were first presented by
\sca{Segal} \cite{Seg} in a form equivalent to the above.
The quantum space of $KL[\Cz^n]$ was systematically studied by
\sca{Bargmann} \cite{Bar}.

\bigskip
{\bf Example: The Bloch sphere.}
The unit sphere in $\Cz^2$ is a coherent manifold $Z_{2j+1}$ with
coherent product $K(z,z'):=(z^*z')^{2j}$ for some
$j=0,\half,1,\frac{3}{2},\dots$.
It corresponds to the Poincar\'e sphere (or Bloch sphere)
representing a single quantum mode of an atom with spin $j$,
or for $j=1$ the polarization of a single photon mode.
The corresponding quantum space has dimension $2j+1$. The associated
coherent states are the so-called \bfi{spin coherent states}.

This example shows that the same set $Z$ may carry many
interesting coherent products, resulting in different coherent spaces
with nonisomorphic quantum spaces.

\bigskip
{\bf Example: The classical limit.}
In the limit $j\to\infty$, the unit sphere turns into
the coherent space of a classical spin, with coherent product
\[
K(z,z'):=\cases{1 & if $z'=\ol z$,\cr 0 & otherwise.}
\]
The resulting quantum space is infinite-dimensional and describes
\bfi{classical stochastic motion} on the Bloch sphere in the Koopman
representation.

More generally, any coherent space $Z$ gives rise to an infinite
family of coherent spaces $Z_n$ on the same set $Z$ but with modified
coherent product $K_n(z,z'):=K(z,z')^n$ with a nonnegative
integer $n$. (The need for a nonnegative integer is related to
Bohr--Sommerfeld quantization.)
The quantum space $\Qz(Z_n)$ is the symmetric tensor product of $n$
copies of the quantum space $\Qz(Z)$.
If $Z$ has a physical interpretation and the classical limit
$n\to\infty$ exists, it usually has a physical meaning, too.

The same abstract quantum system may allow different classical views.
The most conspicuous expression of this ambiguity is the
\bfi{particle-wave duality}, a notion describing the seemingly
paradoxical situation that the same quantum system may be
approximately interpreted either in terms of classical particles or in
terms of classical waves, though depending on the circumstances only
one of the approximate views may be accurate enough to be useful.
This is accommodated by writing the same Hilbert space in different but
isomorphic ways as the quantum space of different coherent spaces.

Complex quantum systems with finitely many degrees of freedom can
be modeled on the same phase spaces as the corresponding classical
systems, and with little additional conceptual effort. (Traditionally,
one would need the second quantization formalism or a first quantized
equivalent.)
The possibility to describe \bfi{motion} is added by augmenting the
state space by variables for position and momentum.
Several particles are accounted for by taking the direct product of the
single-particle phase spaces, the coherent products simply multiply.

\bigskip
{\bf Example: Subsets of a Euclidean space.}
Any subset $Z$ of a Euclidean space $\Hz$ is a coherent space with
coherent product $K(z,z'):=z^*z'$. If the linear combinations of $Z$ 
are dense in $\Hz$, then $\ol\Qz(Z)=\ol \Hz$.
Conversely, any coherent space arises in this way from its quantum
space.

\bigskip
{\bf Example: Quantum spaces of entire functions.}
A \bfi{de Branges function} (\sca{de Branges}  \cite{deBra})
is an entire analytic function $E:\Cz\to\Cz$ satisfying
\[
|E(\ol z)|<|E(z)| \iif \im z>0.
\]
With the  coherent product
\[
K(z,z'):=\cases{
  \ol E'(\ol z)E(z')-E'(\ol z)\ol E(z')   & if $z'=\ol z$,\cr
                                 ~~~      &~\cr
\D\frac{\ol E(\ol z)E(z')-E(\ol z)\ol E(z')}{2i(\ol z-z')} & otherwise,}
\]
where $E'(z)$ denotes the derivative of $E(z)$ with respect to $z$,
$Z=\Cz$ is a coherent space.
The corresponding quantum spaces are the de Brange spaces relevant in
complex analysis.

\bigskip
Coherent spaces and reproducing kernel Hilbert spaces are mathematically
almost equivalent concepts, and there is a vast literature related to
 the latter. Most relevant for the present work are the books by
\sca{Perelomov} \cite{Per} and \sca{Neeb} \cite{Nee}; for applications
in probability and statistics, see also \sca{Berlinet \& Thomas-Agnan}
\cite{BerT}.

However, the emphasis in these books is quite different from the
present exposition, as they are primarily interested in properties of
the associated functions, while we are primarily interested in the
geometry and symmetry properties and in computational tractability.

\subsection{New states from old ones}\label{ss.newOld}

From the set of coherent states it is possible to create a large number 
of other states whose inner product is computable by a closed formula.
This is important for numerical applications, since one can pick from
the new states created in this fashion a suitable subset and declare the
states belonging to this subset to be the coherent states of a new, 
derived coherent space. This way of constructing new coherent spaces 
from old ones allows one to apply the general body of techniques for the
analysis of coherent spaces and their quantum properties to the new
coherent space. Many known numerical techniques for quantum physics 
problems become in this way organized in the same setting.

The first, often useful construction takes a path $u(t)$ in $Z$ and
creates new states
\[
[R_tu(t)]:=\lim_{h\downto 0} h^{-1}\Big(|u(t+h)\>-|u(t)\>\Big).
\]
We write $=\partial_j K$ for the partial derivative with respect to 
the $j$th argument of $K$, and find the inner products 
\[
\<z|[R_tu(t)]=\lim_{h\downto 0} h^{-1}\Big(K(z,u(t+h))-K(z,u(t))\Big)
=\partial_2K(z,u(t))\dot u(t),
\]
\[
\bary{lll}
[R_tu(t)]^*[R_sv(s)]
&=&\D\lim_{h\downto 0} 
h^{-1}\Big(\<u(t+h)|[R_sv(s)]-\<u(t)|[R_sv(s)]\Big)
\\
&=&\D\lim_{h\downto 0} 
h^{-1}\Big(\partial_2K(u(t+h),v(s))\ol{\dot v(t)}
          -\partial_2K(u(t),v(s))\ol{\dot v(t)}\Big)
\\[4mm]
&=&\dot u(t)\partial_1\partial_2K(u(t),v(s))\dot v(s).
\eary
\]
Similar expressions can be found by taking 
other smooth parameterizations of submanifolds of $Z$, and taking limits
corresponding to first order or higher order partial derivatives.

A trivial construction is to take linear combinations 
\[
[\alpha,y]:=\sum_k\alpha_k|y_k\>,
\]
where $\alpha$ is a finite sequence of complex numbers $\alpha_k$
and $y$ is a finite sequence of points $y_k\in Z$.
The inner products are given by
\[
\<z|[\alpha,y]=\sum_k\alpha_k K(z,y_k),
\]
\[
[\alpha,y]^*[\alpha',y]'=\sum_{j,k}\alpha_j\alpha_k'K(y_j,y_k),
\]
This also works for infinite sequences provided the right hand sides
are always absolutely convergent, and with sums replaced by integrals 
for weighted integrals $\int\alpha(x)|y(x)\>d\mu(x)$, provided the
corresponding integrals on the right hand sides are always absolutely 
convergent. Of course, all these recipes can also be combined.

We see that, unlike in traditional Hilbert spaces, where the 
calculation of inner products always requires to evaluate often 
high-dimensional integrals, here the calculation of inner products
is much simpler, often only taking sums and derivatives.

\section{Coherent spaces and quantization}\label{s.cohQuant}

This section introduces the concept of symmetries (invertible coherent 
maps) of coherent spaces and associated quantization procedures. 
This leads to quantum dynamics, which in special (completely integrable)
situations can be solved in closed form in terms of classical motions 
on the underlying coherent space, if the latter has a compatible 
manifold structure. Spectral issues can in favorable cases be handled 
in terms of dynamical Lie algebras. Close relations to concepts from 
geometric quantization and K\"ahler manifolds are pointed out.

\subsection{Symmetries}\label{ss.sym}

Symmetries of a coherent space are transformations of the space that
preserve the coherent structure. They generalize canonical
transformations of a symplectic manifold, which is the special case of
classical mechanics of point particles. More specifically, a
\bfi{symmetry} of a coherent space $Z$ is a bijection $A$ of $Z$ with
the property that
\lbeq{e.sym}
K(z,Az')=K(A^Tz,z')
\eeq
for another bijection $A^T$.

Let $Z$ be a coherent space.
A map $A:Z\to Z$ is called \bfi{coherent} if there is an
\bfi{adjoint map} $A^*:Z\to Z$ such that
\lbeq{e.cohadj}
K(z,Az')=K(A^*z,z') \for z,z'\in Z
\eeq
If $Z$ is nondegenerate, the adjoint is unique, but not in general.

A \bfi{symmetry} of $Z$ is an invertible coherent map on $Z$ with an
invertible adjoint.

Coherent maps form a semigroup $\coh Z$ with identity; the symmetries
form a group.

An \bfi{isometry} is a coherent map $A$ that has an adjoint satisfying
$A^*A=1$. An invertible isometry is called \bfi{unitary}.

Symmetries of a coherent space often represent the dynamical symmetries
(see, e.g., \sca{barut \& Raczka} \cite{BarR}) of an
associated exactly solvable classical system. For example, if $Z$ is a
line bundle over a symplectic phase space, the symmetries would be all
linear symplectic maps and their central extensions. (But only some of
them preserve the Hamiltonian and hence are symmetries of the system
with this Hamiltonian.)

In the coherent space formed by a subset $Z$ of
$\Cz^n$ closed under conjugation, with coherent product
$K(z,z'):=z^Tz'$, all $n\times n$ matrices mapping $Z$ into itself are
(in this particular case linear) coherent maps, and all invertible
matrices are symmetries.

\bigskip
{\bf Example: Distance regular graphs.}
The orbits of groups of linear self-mappings of a Euclidean space define
coherent spaces with predefined transitive symmetry groups.
For example, the symmetric group $\Sym(5)$ acts as a group of
Euclidean isometries on the 12 points of the icosahedron in $\Rz^3$.
The coherent space consisting of these 12 points with the induced
coherent product therefore has $\Sym(5)$ as a group of unitary
symmetries. The quantum space is $\Cz^3$.
The skeleton of the icosahedron is a distance-regular graph,
here a double cover of the complete graph on six vertices.
Many more interesting examples of finite coherent spaces are related
to Euclidean representations of distance regular graphs and other
highly symmetric combinatorial objects. See, e,.g., \sca{Brouwer} et al.
\cite{BroCN}.

The importance of coherent maps stems from the fact that there is a
\bfi{quantization operator} $\Gamma$ that associates with every
coherent map $A$ a linear operator $\Gamma(A)$ on the quantum space
$\Qz(Z)$. In the literature, when applied to the special case where
$\Qz(Z)$ is a Fock space, $\Gamma(A)$ is called the \bfi{second
quantization} of $A$.

\bigskip
{\bf Quantization Theorem.}
{\it Let $Z$ be a coherent space and $\Qz(Z)$ a  quantum space
of $Z$. Then for any coherent map $A$ on $Z$, there is a
unique linear map $\Gamma(A):\Qz(Z)\to\Qz(Z)$ such that
\lbeq{e.move}
\Gamma(A)|z\>=|Az\>\Forall z\in Z.
\eeq
We call $\Gamma(A)$ the \bfi{quantization} of $A$ and $\Gamma$ the
\bfi{quantization map}.}

The quantization map furnishes a representation of the semigroup of
coherent maps on $Z$ (and hence of the symmetry group) on the quantum
space of $Z$.
In particular, this gives a \bfi{unitary representation} of the group
of unitary coherent maps on $Z$.

The quantization operator is important as it reduces many computations
with coherent operators in the quantum space of $Z$ to computations in
the coherent space $Z$ itself. By the quantization theorem, large
semigroups of coherent maps produce large semigroups of coherent
operators, which may make complex calculations much more tractable.
Coherent spaces with many coherent maps are often associated with
symmetric spaces in the sense of differential geometry.

This essentially means that symmetries
are those invertible linear transformations of the quantum space that
map coherent states into coherent states, but is expressed without
reference to the quantum space. This has very important implications
for practical computations, reducing computations in the quantum space
to simple computations in the coherent space. In particular, this makes
certain problems easily exactly solvable that are in the traditional
position or momentum representations nearly intractable.
For example, the calculation of q-expectations requires in the
traditional setting the evaluation of an integral over configuration
space. In case of field theory, the configuration space is
infinite-dimensional, and already a rigorous definition of such
integrals is very difficult. Moreover, finding  closed formulas for
integrals in high or infinite dimensions is more an art than a
science. In contrast, in the coherent space approach, many q-expectations
of interest can be obtained by differentiation, which is a fully
algorithmic process.

In case of the trivial coherent product \gzit{e.trivial}, equation
\gzit{e.sym} holds for every $n\times n$ matrix with the usual matrix
transpose. This motivates the general case, and shows in particular
that the trivial coherent space has the \bfi{general linear group}
$GL(n,\Cz)$ of invertible complex $n\times n$ matrices as its group of
symmetries. For virtually all quantum systems of interest there is a
large classical \bfi{dynamical symmetry group}, which describes the
symmetries of the underlying coherent space. Typically, this symmetry
group is a (possibly infinite-dimensional) Lie group, much larger than
the symmetry group of the system itself -- which is the subgroup
commuting with the Hamiltonian (in the nonrelativistic case) or
preserving the action (in the relativistic case).

\bigskip
{\bf Example: M\"obius space.}
The \bfi{M\"obius space} $Z=\{z\in\Cz^2 \mid |z_1|>|z_2|\}$ is a
coherent manifold with coherent product
$K(z,z'):=(\ol z_1z_1'-\ol z_2z_2')^{-1}$.
A quantum space is the Hardy space of analytic functions on the complex
upper half plane with Lebesgue integrable limit on the real line.
The M\"obius space has a large semigroup of coherent maps (a
semigroup of compressions, \sca{Olshanski \cite{Ols}}) consisting of the
matrices $A\in\Cz^{2\times 2}$ such that
\[
\alpha:=|A_{11}|^2-|A_{21}|^2,~~~
\beta:=\ol A_{11}A_{12}-\ol A_{21}A_{22},~~~
\gamma:=|A_{22}|^2-|A_{12}|^2
\]
satisfy the inequalities
\[
\alpha>0,~~~|\beta|\le \alpha,~~~\gamma\le \alpha-2|\beta|.
\]
It contains as a group of symmetries the group $GU(1,1)$ of matrices
preserving the Hermitian form $|z_1|^2-|z_2|^2$ up to a positive factor.

\bigskip
{\bf Highest weight representations.}
The example of the  M\"obius space generalizes to a large class of 
exactly solvable classical systems with finitely many degrees of 
freedom, corresponding to the coherent states from group 
representations discussed in \sca{Zhang} et al. \cite{ZhaFG} and 
\sca{Simon} \cite{Sim}, which are close to being 
computable (though not all needed details are in these papers).
The constructions relate to central extensions of all
\bfi{semisimple Lie groups} and associated \bfi{symmetric spaces}
or \bfi{symmetric cones} and their \bfi{line bundles}.
These provide many interesting examples of coherent manifolds.
This follows from work on coherent states constructed from highest
weight representations, discussed in monographs by
\sca{Perelomov} \cite{Per}, by \sca{Faraut \& Kor\'anyi} \cite{FarK},
by \sca{Neeb} \cite{Nee}.

Coherent states from highest weight
representations induce on the corresponding coadjoint orbit a measure,
a metric, a symplectic form, and an associated symplectic Poisson
bracket. (See the \sca{Zhang} et al. \cite{ZhaFG} survey for
details from a physical point of view. The Poisson bracket defines a
Lie algebra on phase space functions ($C^\infty$ functions on the
coadjoint orbit,
hence an associated group of Hamiltonian diffeomorphisms, and the
coherent state approach effectively quantizes this group. All this can
be reconstructed directly from the associated coherent spaces.
In particular, the nonclassical states of light in quantum optics
called \bfi{squeezed states} are described by coherent spaces
corresponding to the \bfi{metaplectic group}; cf. related work by
\sca{Neretin} \cite{Ner}.

\subsection{q-observables and dynamics}

We now assume that  $Z$ is a \bfi{coherent manifold}. This means that
$Z$ carries a $C^\infty$-manifold structure with respect to which the
coherent product is smooth ($C^\infty$).
The relevant \bfi{observables} of the classical system are
the discrete symmetries and the infinitesimal generators of the
1-parameter groups of symmetries that are smooth on the coherent
product. They are promoted to q-observables
of the corresponding quantum system through the quantization map.
For a symmetry $A$, the corresponding q-observable is $\Gamma(A)$.
For an \bfi{infinitesimal symmetry} $X$, i.e., an element of the Lie 
algebra of generators of 1-parameter groups of the symmetry group),
the corresponding \bfi{quantum symmetry}, acting on the quantum space 
of $Z$, is the q-observable given by the strong limit
\[
 d\Gamma(X):=\lim_{s\downto 0}\frac{\Gamma(e^{isX})-1}{is}.
\]
Note that
\[
 d\Gamma(X+Y)= d\Gamma(X)+ d\Gamma(Y),~~~
e^{ d\Gamma(X)}=\Gamma(e^X).
\]
The quantization theorem from Subsection \ref{ss.sym} may be regarded 
as a generalized \bfi{Noether principle}
that automatically promotes all symmetries of $Z$ to dynamical
symmetries of the corresponding quantum system.

Thus a coherent space contains intrinsically all information needed to
interpret the quantum system, including that about which operators may
be treated as q-observables.

The dynamics of a physical system is traditionally given by a
\bfi{Hamiltonian}, a symmetric and Hermitian expression $H$ in the
q-observables.
If the coherent space is in fact a \bfi{coherent manifold}, the
\bfi{classical dynamics} determined by the Hamiltonian is given by a
Poisson bracket canonically associated to the coherent space through
variation of the so-called \bfi{Dirac--Frenkel action} discussed 
in Subsection \ref{ss.var} below.
(Classical mechanics on Poisson manifolds, the most general setting
for the dynamics in closed classical systems, is discussed in detail in
\sca{Marsden \& Ratiu} \cite{MarR}. Less general is classical mechanics
on symplectic manifolds, and even more restricted is classical mechanics
on cotangent bundles, which includes classical mechanics on phase space
$\Rz^{6N}$ for systems of $N$ particles in Cartesian coordinates.)

In a classical Hamiltonian system, the dynamics of a phase space
function $f$ is given by $\dot f=H\lp f$ where $f\lp g=\{g,f\}$ in
terms of the Poisson bracket. For an $N$-particle system with particle
positions $q_j$ and particle momenta $p_j$, specializing this to
$f=q_j$ and $g=p_j$ gives the classical equations of motion.
In a quantum system, one has the same in the Heisenberg picture, and
according to every textbook, the resulting dynamics is equivalent to
the Schr\"odinger equation in the Schr\"odinger picture.

{\bf Exactly solvable systems.}
In the special case where the classical Hamiltonian is an infinitesimal 
symmetry of $Z$, and hence the quantum Hamiltonian has the form 
$\Gamma(H)$, the quantization lifts
the classical phase space trajectory to a quantum trajectory.
Thus if the Lie algebra of q-observables contains the Hamiltonian
(and in some slightly more general situations),
the quantum dynamics has the special feature that coherence is
dynamically preserved. 
In terms of the Hamiltonian, the dynamics for pure quantum states $\psi$
is traditionally given by the \bfi{time-dependent Schr\"odinger 
equation}
\[
i\hbar \frac{d\psi}{dt} =  d\Gamma(H)\psi
\]
for the corresponding \bfi{quantum Hamiltonian} $ d\Gamma(H)$.
A dynamical symmetry preserved by $H$ (in the classical case) or
$ d\Gamma(H)$ (in the quantum case) is a true symmetry of the
corresponding classical or quantum system.
The Fourier transform $\wh \psi(E)$ satisfies the \bfi{time-independent
Schr\"odinger equation}
\lbeq{e.tiSchr}
 d\Gamma(H)\wh \psi=E\wh \psi.
\eeq
The following result shows that the solution of Schr\"odinger equations
with a sufficiently nice Hamiltonian can be reduced to solving
differential equations on $Z$.

\begin{thm}
Let $Z$ be a coherent space, and $\Gz$ be a Lie group of coherent maps
with associated Lie algebra $\Lz$.
Let $H(t)\in\Lz$ be a Hamiltonian with possibly time-dependent
coefficients. Then the solution of the initial value problem
\lbeq{eg53}
  i \hbar \D \frac{\partial }{\partial t} \psi_t =
   d\Gamma(H(t)) \psi_t,~~\psi_0=|z_0\>
\eeq
with $z_0\in Z$
has for all times $t \ge 0$ the form of a coherent state,
$\psi_t=|z(t)\>$ with the trajectory $z(t)\in Z$ defined by the
initial value problem
\[
i\hbar\dot z(t)=H(t)z(t),~~~z(0)=z_0.
\]
\end{thm}

This means that if a system is at some time in
a coherent state it will be at all times in a coherent state.

This conservation of coherence has the consequence that the quantum
system is \bfi{exactly solvable}. This means that the complete solution
of the dynamics of the quantum system can be reduced to the solution of
the corresponding classical system. Effectively, the partial
differential equations of quantum mechanics in the quantum space are
solved in terms of ordinary differential equations on the underlying
coherent space. In many cases, this implies that the spectrum can be
determined explicitly in terms of the representation theory of the
corresponding Lie algebras.

More generally (see, e.g., \sca{Iachello} \cite{Iac}), we have an 
exactly solvable system whenever the Hamiltonian $H(t)$ is a linear 
combination of infinitesimal symmetries with coefficents given by 
\bfi{Casimirs} of the Lie algebra $\Lz$ of infinitesimal symmetries, 
i.e., in the classical case central elements of the Lie--Poisson 
algebra $C^\infty(\Lz^*)$, and in the quantum case of the universal 
enveloping algebras of $\Lz$. On any orbit of the symmetry group, these 
Casimirs are represented by multiplication with a constant. One can 
therefore extend the coherent space $Z$ without changing the quantum 
space by treating the corresponding multiples of the coherent states 
as new coherent states of an extended coherent space whose 
elements are labelled by pairs of elements of $Z$ and appropriate 
multipliers. This turns the algebra of Casimirs into an abelian group 
of symmetries of the extended coherent space, which, together with 
original symmetries provides an action of a central extension of the
original symmetry group as a symmetry group of the extended coherent 
space.

The time-independent Schr\"odinger equation \gzit{e.tiSchr} generalizes
easily to a more general \bfi{implicit Schr\"odinger equation}
$I(E)\psi=0$. This more general formulation fits naturally the coherent 
space setting, and everything said so far (corresponding to 
$I(E)= E-d\Gamma(H)$) generalizes to the general implicit formulation.

\subsection{Dynamical Lie algebras}\label{ss.dynLie}

A quantum dynamical problem can often be reduced to finding the 
spectrum of a physical system defined by an implicit Schr\"odinger 
equation 
\lbeq{e.ISE}
I(E)\psi=0
\eeq
with an energy-dependent \bfi{system operator} $I(E)$, and $\psi$ in 
the antidual of some Euclidean space $\Hz$.
A nonlinear $I(E)$ typically appears in reduced effective descriptions 
of systems derived from a more complicated Hamiltonian setting and in 
relativistic systems. (The antidual is needed to account for a possible 
continuous spectrum.) 

This section discusses implicit Schr\"odinger equations for the exactly 
solvable case where the system operator $I(E)$ is contained in a 
Lie algebra $\Lz$ with known representation theory. This is the setting
where a tractable dynamical symmetry group for the Hamiltonian is known
and covers many interesting systems.

For example, the system operator $I(E)=p_0^2-\p^2-(mc)^2$ with $p_0=E/c$
describes a free spin 0 particle. This generalizes to a quadratic 
implicit Schr\"odinger equation
\lbeq{e.ISEspin}
\Big(\pi^2 - \frac{ige \hbar}{c} \SS\cdot\F(x) -(mc)^2\Big)\psi=0
\eeq
for a particle of charge $e$, mass $m$, and arbitrary spin in an 
electromagnetic field. Here
\lbeq{e.pi}
\pi={\pi_0 \choose\ppi} := p + e\hbar A(x)
\eeq
is a gauge invariant 4-vector, $\SS$ is the 3-dimensional spin vector 
representing the intrinsic angular momentum of a particle of spin
 $j=0,\half,1,\ldots$, the 3-vector $\F(x)=\E(x)+ic\B(x)$ is the 
Riemann--Silberstein vector encoding the electric field $\E(x)$ and the 
magnetic field $\B(x)$, and $g$ is the dimensionless \bfi{g-factor} of 
the \bfi{magnetic moment}
\[
\mmu_s:=-\frac{g\mu_B}{\hbar}\SS,
\]
where $\mu_B$ is a constant called the \bfi{Bohr magneton}, and $\psi$ 
is a wave function with $s=2j+1$ components. For spin $j=1/2$, we have 
$s=2$ components, hence, being second order, 4 local degrees of freedom,
corresponding to the 4 components of the (first order) Dirac equation, 
which is equivalent to the special case $g=2$.

In the special case where the dependence on $E$ is linear, we have
\lbeq{e.linear}
I(E)=EM-N
\eeq
with fixed $M,N\in\Lz$. This covers the simple case of a harmonic 
oscillator, where $M=1$, $N=\half(p^2/m + Kq^2)$ is the Hamiltonian, 
and the Lie algebra is the \bfi{oscillator algebra}, with generators 
$1, p,q, H$ (or, in complex form, $1, a, a^*, a^*a$).
It also covers a family of practically relevant exactly solvable systems
with  Lie algebra $\Lz=so(2,1)\oplus\Cz=su(1,1)\oplus\Cz$ discussed in 
detail in the book by \sca{Wybourne} \cite{Wyb}, containing among others
the case of a particle of mass $m$ in a Coulomb field, with $~M=r=|\q|$
and $N=MH$, where 
\[
H=\half m\vv^2-\frac{\alpha}{|\q|}
\] 
is the Coulomb Hamiltonian.

If $I(E)$ belongs for all $E$ to some Lie algebra $\Lz$ acting on 
$\Hz$ in a (reducible or irreducible) representation then $\Lz$ is 
called a \bfi{dynamical Lie algebra}\footnote{
One can always take the dynamical Lie algebra to be the Lie algebra 
$\Lin\Hz$ of all linear operators on the nuclear space $\Hz$. 
For this choice, the dynamical Lie algebra offers no advantage over the 
standard treatment. 
Therefore it is usually understood that the dynamical Lie algebra is 
much smaller than $\Lin\Hz$, although mathematically there is no such
restriction.
}
of the problem.

In general, the requirement for a dynamical symmetry group is just
that all quantities of physical interest in the system can be 
expressed in the Lie--Poisson algebra (in the classical case) or the 
universal enveloping algebra (in the quantum case) of the corresponding 
Lie algebra. In this case, the label ``dynamical'' is a misnomer, and 
\bfi{kinematic symmetry group} would be more appropriate. The kinematic
symmetry group is an integral part of the Hamiltonian or Lagrangian 
setting; so one usually gets it directly from the formulation and a 
look at the obvious symmetries. For any anharmonic oscillator it is 
$Sp(2)$; for any system of $N$ particles in $\Rz^3$ it is the 
symplectic group $Sp(6N)$, generated by the inhomogeneous quadratics 
in $p$ and $q$.

If a problem has a dynamical symmetry group such that the (discrete or 
continuous) spectrum of all elements of its Lie algebra $\Lz$ is 
exactly computable then the spectrum of the system can be found exactly.
In the best understood cases, $\Lz$ is a finite-dimensional semisimple 
Lie algebra. Here everything is tractable more or less explicitly since
the representation theory of these Lie algebras and their corresponding 
groups is fully understood.
A problem solvable in this way is called \bfi{integrable}. 

The \bfi{spectrum} of the nonlinear eigenvalue problem \gzit{e.ISE} is 
the set $\spec I$ of all $E\in\Cz$ such that $I(E)$ is not invertible.
In terms of (generalized) eigenvalues and eigenvectors of $I(E)$,
\[
I(E)|\xi,E\>=\lambda(\xi,E)|\xi,E\>,
\]
where $\xi$ is a label distinguishing different eigenvectors $|\xi,E\>$
in a (generalized) orthonormal basis of the eigenspace corresponding to 
the eigenvalue $E$. To cover the continuous spectrum (where eigenvectors
are unnormalized, hence do not belong to the Hilbert space), we work in 
a Euclidean space $\Hz$ on which the Hamiltonian acts as a linear 
operator. The Hilbert space of the problem is then the completion 
$\ol\Hz$ of this space, and $ \Hz \subseteq \ol\Hz \subseteq \Hz^\times$
is a Gelfand triple. Therefore 
\[
I(E)|\xi,E\>=0 \mbox{~~~whenever~} \lambda(\xi,E)=0.
\]
Thus 
\[
\spec I=
\{E\in\Rz\mid \lambda(\xi,E)=0 \mbox{~for some~} \xi\in\spec I(E)\}\\
\]
Moreover, it is easy to see that all eigenvectors of the nonlinear 
eigenvalue problem have the form $|\xi,E\>$. Thus the spectrum is given 
by the set of solutions of the nonlinear equation $\lambda(\xi,E)=0$.

In many cases of interest (e.g., cf. Subsection \ref{ss.Lie*}, when 
$\Lz$ is a Lie $*$-algebra), $\Lz=\Lz_0\oplus\Cz$; then we may write 
\lbeq{e.isolv2}
I(E):=m(E)X(E)-k(E),
\eeq
where $m(E)$ and $k(E)$ are scalars not vanishing simultaneously, and 
$X(E)\in \Lz_0$. If
\[
X(E)|\xi,E\>=\xi|\xi,E\>
\]
is a complete system of (generalized) eigenvalues and eigenvectors of 
$X(E)$ then 
\lbeq{e.asSpec}
I(E)|\xi,E\>=\lambda(\xi,E)|\xi,E\>,~~~
\lambda(\xi,E)=m(E)\xi-k(E).
\eeq
Therefore 
\[
I(E)|\xi,E\>=0 \mbox{~~~whenever~} \lambda(\xi,E)=0.
\]
Again, all eigenvectors of the nonlinear eigenvalue problem have the 
form $|\xi,E\>$, and the spectrum is given by the set of solutions of 
$\lambda(\xi,E)=0$.

If a problem has a dynamical symmetry group such that the (discrete or 
continuous) spectrum of all elements of its Lie algebra $\Lz$ is 
exactly computable then the spectrum of the system can be found exactly.
In the best understood cases, $\Lz$ is a finite-dimensional semisimple 
Lie algebra. Here everything is tractable more or less explicitly since
the representation theory of these Lie algebras and their corresponding 
groups is fully understood. In this case, one may find the $|s,E\>$
by transforming $I(E)$ to elements from a standard set of 
representatives of the conjugacy classes, and has to work out 
explicit spectral factorizations for these.
For semisimple Lie algebras $\Lz$ in finite dimensions, each Lie 
algebra element is in a Cartan subalgebra, and the latter are all 
unitarily conjugate, i.e., if $V$ and $V'$ are cartan subalgebras, 
there is a group element $U$ such that $V'=\{\ad_U X \mid X\in V\}$. 
So one only has to consider conjugacy inside the standard Cartan 
subalgebra. (In the noncompact case, the eigenvectors correspond to 
representatives from any conjugacy class, which may be several in the 
same irreducible represention)
This is enough to give the spectrum, and in the discrete case the full 
spectral resolution. In the continuous case, one still needs to find 
the spectral density and from it the S-matrix; cf. \sca{Kerimov} 
\cite{Ker1,Ker2}.

\subsection{Relations to geometric quantization} \label{ss.geomQuant}

Often, classical symmetries are promoted to quantum symmetries
in a projective representation. Then the symmetry group of the extended
phase space is a proper central extension of the symmetry group of the 
original space. It acts on an extended phase space whose dimension is l
arger. For example, the classical phase space with $n$ spatial degrees 
of freedom has dimension $2n$, but the associated Heisenberg algebra, 
the central extension of an abelian group with $2n$ generators, has 
dimension $2n+1$, as the canonical commutation relations for the 
extended Poisson bracket (or in the quantum case for the commutator) 
require an additional central generator.

Such a central extension is the rule rather than the exception.
The extra dimension, often called \bfi{Berry phase} or 
\bfi{geometric phase}, accounts for topological features such as the 
Aharonov--Bohm effect.
But it also occurs in classical physics; e.g., a classical
electromagnetic field exhibits topological effects when the field
strength is not globally integrable to a vector potential.

The explicit description of a central extension in terms of the original
symmetry group involves so-called \bfi{cocycles}. Rather than with the 
original symmetry group, one can indeed work directly with a central 
extension of the group, acting on the extended phase space. (Examples
where this works are the M\"obius space and the Klauder spaces 
discussed before.) 
In this way, one can avoid the use of cocycles, as the relevant 
projective representations become ordinary representations of the 
central extension. Thus the extended description generally reflects the 
quantum properties in a more symmetric way than the original coherent 
space.

In our present setting, the extended phase space is modeled by a
projective coherent space.
A \bfi{projective coherent space} is a coherent space with a
\bfi{scalar multiplication} that assigns to each nonzero complex
number $\lambda$ and each $z\in Z$ a point $\lambda z \in Z$ such that\\
(C4)~ $\ol{\lambda z} ~=~  \ol{\lambda} \, \ol z,~~~
\lambda(\mu z)=(\lambda\mu)z$~;\\
(C5)~ $K(\lambda z,z') ~=~ \lambda^e K(z,z')$\\
for some nonzero integer $e$.
Projectivity is typically needed when one wants to have all symmetries
of interest represented coherently.  Projective coherent spaces
coherently
represent central extensions of groups in cases where the original
group is represented by a projective representation that would lead to
coherent maps only up to additional scalar factors called cocycles.

Geometrically, the extended phase space takes the form of a line bundle.
In case of the Heisenberg algebra, the line bundle is trivial, formed by
$Z=\Cz\times\Cz^n$ with componentwise conjugation, scalar
multiplication defined by $\alpha (\lambda,s):= (\alpha \lambda, s)$,
and coherent product
\[
K(z,z'):=\lambda\lambda'e^{s^Ts'/\hbar}
\for z=(\lambda,s),~z'=(\lambda',s')
\]
we get a projective coherent space $Z$ whose quantum space $\Qz(Z)$ is
the bosonic Fock space with $n$ independent oscillators, and the
coherent states are the multiples of the Glauber coherent states.
Indeed, the coherent states
\[
|\lambda, s\>  ~~, ~~~~ \lambda,s \in \Cz
\]
in a single-mode Fock space have the Hermitian inner product
\[
\<\lambda,s \mid \lambda', s'\>
    ~=~ \ol \lambda\lambda '~ e^{\ol s s'/\hbar}.
\]
By means of Klauder spaces (defined above), the construction easily 
extends to an arbitrary finite or infinite number of modes. In terms of 
the traditional Fock space description, the coherent states are the
simultaneous eigenstates of the annihilator operators,
\[
a|z\>=\z|z\> \for z\in Z.
\]
More generally (see \sca{Neumaier \& Ghaani Farashahi}
\cite{NeuF.cohQuant}), Klauder spaces provide an elegant and efficient 
approach to the properties of creation and annihilation operators. 

A coherent space generalizes finite-dimensional symplectic manifolds
with a polarization that induces a complex structure on the
manifold. A projective coherent space generalizes a corresponding
\bfi{Hermitian line bundle} $Z$, i.e., a line bundle with a Hermitian
connection. Such line bundles are 
usually discussed in the context of geometric quantization.

\bfi{Geometric quantization} (see, e.g., 
\sca{Bar-Moshe \& Marinov} \cite{BarMosM},
\sca{Engli{\v s}} \cite{Eng96,Eng}, \sca{Schlichenmaier} \cite{Schli})
proceeds from a symplectic manifold $\Mz$. It constructs
(in the group case in terms of integral\footnote{
Integral cohomology apparently corresponds to the fact that the line
bundle can in fact be viewed as a $U(1)$-bundle so that phases are
well-defined.
} 
cohomology) a polarization that defines a Hermitian line bundle
$Z = \Cz\Mz$ and an associated \bfi{K\"ahler potential} (which is
essentially the logarithm of the coherent product). This potential
turns $\Mz$ into a K\"ahler manifold with a natural K\"ahler metric,
K\"ahler measure, and symplectic K\"ahler bracket.
If the K\"ahler metric is definite (which is always the case if $Z$ is
a compact symmetric space), there is an associated Hilbert space
of square integrable functions on which quantized operators can be
defined by a recipe of \sca{van Hove} \cite{vHov}.

An \bfi{involutive coherent manifold} is a coherent manifold $Z$
equipped with a smooth mapping that assigns to every $z\in Z$ a
\bfi{conjugate} $\bar z\in Z$ such that $\ol{\ol z} = z$ and
$\ol{K(z,z')}=K(\ol z,\ol z')$ for $z,z'\in Z$. Under additional 
conditions, an involutive coherent manifold carries a canonical 
K\"ahler structure turning it into a \bfi{K\"ahler manifold}.
For semisimple finite-dimensional Lie algebras, the irreducible highest
weight representations have nice coherent space formulations. In the
literature, the logarithm of the coherent product figures under the name
of K\"ahler potential. \sca{Zhang}  et al. \cite{ZhaFG} relate the
latter to coherent states.
The coherent quantization of K\"ahler manifolds is equivalent to
traditional \bfi{geometric quantization of K\"ahler manifolds}.
But in the coherent setting, quantization is not restricted to
finite-dimensional manifolds, which is important for quantum field
theory.

The coherent product and the
conjugation are $C^\infty$-maps on the line bundle. In order that
this line bundle exists, the symplectic manifold must also carry a
positive definite K\"ahler potential $F:Z\times Z\to\Cz$
satisfying a generalized \bfi{Bohr--Sommerfeld quantization condition}
defined by the integrality of some cohomological expression.
In this case, $Z$ is a projective coherent space with coherent product
\[
K(z,z'):=e^{-F(z,z')}.
\]
The quantum space of the projective coherent space carries the
representation satisfying the conditions of a successful geometric
quantization.
This procedure, called \bfi{Berezin quantization}, is the most useful
way of performing geometric quantization; see, e.g.,
\sca{Schlichenmaier} \cite{Schli}.

\section{The thermal interpretation in terms of Lie algebras}
\label{s.TILie}

Coherent quantum physics on a coherent space $Z$ is related to physical 
reality by means of the thermal interpretation, discussed in detail in 
Part II \cite{Neu.IIfound} and applied to measurement in Part III 
\cite{Neu.IIIfound} of this series of papers. We rephrase the formal 
essentials of the thermal interpretation in a slightly generalized more 
abstract setting, to emphasize the essential mathematical features and 
the close analogy between classical and quantum physics. We show how
the coherent variational principle (the Dirac--Frenkel procedure 
applied to coherent states) can be used to show that in coarse-grained 
approximations that only track a number of relevant variables, quantum 
mehcnaics exhibits chaotic behavior that, according to the thermal 
interpretation, is responsible for the probabilistic features of 
quantum mechanics.

\subsection{Lie $*$-algebras}\label{ss.Lie*}

A (complex) \bfi{Lie algebra} is a complex vector space $\Lz$ with a 
distinguished \bfi{Lie product}, a bilinear operation\index{$\lp$} on 
$\Lz$ satisfying $X\lp X = 0$ for $X\in\Lz$ and the 
\bfi{Jacobi identity}
\[
X\lp (Y\lp Z) + Y \lp (Z \lp X)+ Z \lp (X \lp Y)=0
\for X,Y,Z\in \Lz.
\]
A \bfi{Lie $*$-algebra} is a complex Lie algebra $\Lz$ with a 
distinguished element $1\ne 0$ called \bfi{one} and a mapping
$*$ that assigns to every $X\in\Lz$ an {\bfi{adjoint}} $X^*\in\Lz$
such that 
\[
(X+Y)^*= X^*+Y^*,~~~(X\lp Y)^* = X^*\lp Y^*,
\]
\[
X^{**}=X,~~~(\lambda X)^* = \lambda^* X^*,
\]
\[
1^*=1,~~~X\lp 1 =0
\]
for all $X,Y\in \Lz$ and $\lambda\in \Cz$ with complex conjugate 
$\lambda^*$. We identify the multiples of 1 with the corresponding 
complex numbers.

A \bfi{state} on a Lie $*$-algebra $\Lz$ is a positive semidefinite 
Hermitian form $\<\cdot,\cdot\>$, antilinear in the first argument and 
normalized such that $\<1,1\>=1$.

A group $G$ \bfi{acts} on a Lie $*$-algebra $\Lz$ if for every $A\in G$,
there is a linear mapping that maps $X\in\Lz$ to $X^A\in\Lz$ such that
\[
(X\lp Y)^A =X^A\lp Y^A,
\]
\[
(X^A)^B=X^{AB},~~~(X^A)^*=(X^*)^A,~~~X^1=X,~~~1^A=1
\]
for all $X,Y\in\Lz$ and all $A,B\in G$. Thus the mappings $X\to X^A$ are
$*$-automorphisms of the Lie $*$-algebra. Such a family of mappings is 
called a \bfi{unitary representation} of $G$ on $\Lz$. 

Often, unitary representations arise by writing the Lie $*$-algebra 
$\Lz$ as a vector space of complex $n\times n$ matrices closed under 
conjugate transposition ${}^*$ and commutation, with 
$X\lp Y:=\frac{i}{\hbar}[X,Y]$, and $G$ as a group of unitary 
$n\times n$ matrices such that $X^A:=A^{-1}XA\in\Lz$ for all $X\in\Lz$.

\subsection{Quantities, states, uncertainty}

In classical and quantum physics, \bfi{physical systems} are modeled by 
appropriate Lie $*$-algebras $\Lz$, whose elements are interpreted as 
the \bfi{quantities} of the system modeled. Each physical system may 
exist in different \bfi{instances}; each instance specifies a 
particular system under particular conditions. A state defines the 
\bfi{properties} of an instance of a physical system described by a 
model, and hence what \bfi{exists} in the system. Properties depend on 
the state and are expressed in terms of definite but uncertain values 
of the quantities:

\bfi{(GUP)} \bfi{General uncertainty principle:}
{\it In a given state, any quantity $X\in\Lz$ has the \bfi{uncertain 
value} 
\lbeq{e.Xbar}
\ol X:=\<X\>:=\<1,X\>
\eeq
with an \bfi{uncertainty} of\ \footnote{
Since the state is positive semidefinite, the first expression shows
that $\sigma_X$ is a nonnegative real number. The equivalence of both 
expressions defining $\sigma_X$ follows from $\<X\>=\ol X$ and 
\[
\<X-\ol X,X-\ol X\>=\<X,X\>-\<X,\ol X\>-\<\ol X,X\>+\<\ol X,\ol X\>
=\<X,X\>-\<X\>^*\ol X-\ol X^*\<X\>+|\ol X|^2=\<X,X\>-|\ol X|^2.
\]
} 
} 
\lbeq{e.sigmaX}
\sigma_X:=\sqrt{\<X-\ol X,X-\ol X\>}=\sqrt{\<X,X\>-|\ol X|^2}.
\eeq
Through \gzit{e.Xbar}, each state induces an element $\<\cdot\>$ of the 
\bfi{dual} of $\Lz$, the space $\Lz^*$ of linear functionals on $\Lz$. 

As discussed in Part I \cite{Neu.Ifound} (and exemplified in more detail
in Part III \cite{Neu.IIIfound} and Part IV \cite{Neu.IVfound}), the 
\bfi{interpretation}, i.e., the identification of formal properties 
given by uncertain values with real life properties of a physical 
system, is done by means of 

\bfi{(CC)} \bfi{Callen's criterion} (\sca{Callen} \cite[p.15]{Cal}):
{\it Operationally, a system is in a given state if its properties are 
consistently described by the theory for this state.
} 

This is enough to find out in each single case how to approximately 
measure the uncertain value of a quantity of interest, though it may 
require considerable experimental ingenuity to do so with low 
uncertainty. The uncertain value $\ol X$ is considered informative
only when its uncertainty $\sigma_X$ is much less than $|\ol X|$.

As position coordinates are dependent on a convention about the 
coordinate system used, so all system properties are dependent on the 
\bfi{conventions} under which they are viewed. To be objective, these 
conventions must be interconvertible. This is modeled by a group $G$ 
of \bfi{symmetries} acting transitively both on the spacetime manifold 
$M$ considered and on the set $W$ of conventions. We write these 
actions on the left, so that $A\in G$ maps $x\in M$ to $Ax$ and 
$w\in W$ to $Aw$.

To be applicable to a physical system, a representation of $G$ on the 
Lie $*$-algebra $\Lz$ of quantities must be specified. Depending on the 
model, this representation accounts for conservative dynamics and the 
principle of relativity in its nonrelativistic, special relativistic, 
or general relativistic situation. It also caters for the presence of 
internal symmetries of a physical system. Correspondingly, $G$ may be a 
group of matrices, a Heisenberg group, the Galilei group, the Poincar\'e
group, or a group of volume-preserving diffeomorphisms of a spacetime 
manifold $M$. 

A \bfi{particular physical system} in all its views is described by a 
family of states $\<\cdot,\cdot\>_w$ indexed by a convention $w\in W$
satisfying the covariance condition
\lbeq{e.Aw2}
\<X,Y\>_{Aw}=\<X^A,Y^A\>_w 
\eeq
for $X,Y\in\Lz$, $w\in W$, $A\in G$. In particular, uncertain values 
transform as
\lbeq{e.Aw1}
\<X\>_{Aw}=\<X^A\>_w.
\eeq
A \bfi{subsystem} of a particular physical system is defined by 
specifying a Lie $*$-subalgebra and  restricting the family of states 
to this subalgebra.

If (as is commonly done) we work within a fixed affine coordinate system
in a spacetime (homeomorphic to some) $\Rz^d$, the only conditions 
relevant are when and where a system is described; all other conditions 
are handled implicitly by covariance considerations. In this case, $W$
is simply the spacetime $M$, and $G$ is the group of affine translations
$T_z:x\to x+z$ of $M$ by $z$. In this case, 
\[
\ol X(x)=\<X\>_x,~~~\sigma_X(x)=\sqrt{\<X,X\>_x-|\ol X(x)|^2}
\]
define the \bfi{value} $\ol X(x)$ of $X$ at $x$ and its 
\bfi{uncertainty} $\sigma_X(x)$ at $x$, and \gzit{e.Aw2} and 
\gzit{e.Aw1} become
\[
\<X,Y\>_{x+z}=\<X^{Tz},Y^{Tz}\>_x,~~~\<X\>_{x+z}=\<X^{Tz}\>_x.
\]
The value $\ol X(x)$ is (in principle) \bfi{observable} with 
\bfi{resolution} $\delta>0$ if it varies slowly with $x$ and has a 
sufficiently small uncertainty. More precisely, if $\Delta$ denotes the
set of spacetime shifts that are imperceptible in the measurement 
context of interest, observability with resolution $\delta$ requires 
that 
\[
|\ol A(x+h)-\ol A(x)|\le \delta \for h\in \Delta,
\]
\[
\sigma_X(x)\ll |\ol A(x)| +\delta.
\]
We require that the translation group is 
generated by a covariant \bfi{momentum vector} $p\in\Lz^d$ with 
Hermitian components, in the sense that 
\lbeq{e.momentum}
\frac{\partial}{\partial x_\nu} X^{T_x}=p_\nu\lp X
\eeq
for $X\in \Lz$, $x\in M$ and all indices $\nu$. From the covariance 
condition \gzit{e.Aw2}, we conclude that 
\lbeq{e.prodRule}
\frac{\partial}{\partial x_\nu} \<X,Y\>_x
=\<p_\nu\lp X,Y\>_x+\<X,p_\nu\lp Y\>_x.
\eeq
In particular, the uncertain values satisfy the 
\bfi{covariant Ehrenfest equation}
\lbeq{e.covEhrenfest}
\frac{\partial}{\partial x_\nu} \<X\>_x=\<p_\nu\lp X\>_x
\eeq
discussed in a special case in Part II \cite{Neu.IIfound}.

In classical or quantum multiparticle mechanics (as opposed to field 
theory), space and time are treated quite differently, and we are
essentially in the case $d=1$ of the above, where the convention about 
views of system properties is completely specified by the time 
$t\in \Rz$. In this case, the above specializes to
\[
\ol X(t)=\<X\>_t,~~~\sigma_X(t)=\sqrt{\<X,X\>_t-|\ol X(t)|^2}
\]
The time translation group is generated by a Hermitian 
\bfi{Hamiltonian} $H\in\Lz$, and 
\lbeq{e.ham}
\frac{d}{dt} X^{T_t}=H\lp X.
\eeq
\lbeq{e.prodRule1}
\frac{\partial}{\partial t_\nu} \<X,Y\>_t
=\<H\lp X,Y\>_t+\<X,H\lp Y\>_t.
\eeq
In particular, the uncertain values satisfy the \bfi{Ehrenfest equation}
\lbeq{e.Ehrenfest}
\frac{d}{dt} \<X\>_t=\<H\lp X\>_t,
\eeq
providing a deterministic dynamics for the q-expectations.

\subsection{Examples}

1. A simple classical example is $\Lz=\Cz^3$ with the cross product as 
Lie product. It is isomorphic to the Lie algebra $so(3,\Cz)$ and 
describes in this representation a rigid rotator. The dual space 
$\Lz^{*}$ is spanned by the three components of $J$, and 
the functions of $J^2$ are the Casimir operators. Assigning to $J$ a 
particular 3-dimensional vector with real components (since $J$ has 
Hermitian components) gives the classical angular momentum 
in a particular state. 

\bigskip

2. The same Lie algebra is also isomorphic to $su(2)$, the Lie algebra 
of traceless Hermitian $2\times 2$ matrices, and then describes the 
thermal setting of a single qubit. 
In this case, we think of  $\Lz ^{*}$ as mapping the three Hermitian 
Pauli matrices $\sigma_j$ to three real numbers $S_j$, and extending 
the map linearly to the whole Lie algebra. Augmented by $S_0=1$ to 
account for the identity matrix, which extends the Lie algebra to that 
of all Hermitian matrices, this leads to the classical description of 
the qubit discussed in Subsection 3.5 of Part III \cite{Neu.IIIfound}. 

\bigskip

3. Consider the Lie $*$-algebra $\Lz$ of smooth functions $f(p,q)$ on 
classical phase space with the negative Poisson bracket as 
Lie product and $*$ as complex conjugation. Given a $*$-homomorphism 
$\omega$ with respect to the associative pointwise multiplication, 
determined by the classical values $\ol p_k:=\omega(p_k)$ and 
$\ol q_k:=\omega(q_k)$, the states defined by
\[
\<X,Y\>:=\omega(X^*Y)
\]
reproduce classical deterministic dynamics. More generally, $\Lz$ can be
partially ordered by defining $f\ge 0$ iff $f$ takes values in the 
nonnegative reals. Given a monotone $*$-linear functional $\omega$ on 
$\Lz$ satisfying $\omega(X^*)=\omega(X)^*$ for $X\in\Lz$ and 
$\omega(1)=1$, the states defined by
\[
\<X,Y\>:=\omega(X^*Y)
\]
reproduce classical stochastic dynamics in the Koopman picture discussed
in Subsection 4.1 of Part III \cite{Neu.IIIfound}. In both cases, 
$\<X\>=\omega(X)$.

\bigskip

4. The basic example of interest for isolated quantum physics is the Lie
$*$-algebra $\Lz(Z)$ of linear operators acting on the quantum space 
$\Hz=\Qz(Z)$ of the coherent space $Z$, with Lie product
\lbeq{e.QLie}
X\lp B:= \frac{i}{\hbar}[X,B]=\frac{i}{\hbar}(XB-BX).
\eeq
The action of the translation group on $X\in\Lz$ is given by
\[
X^{T_x}:=U(x)^*XU(x)
\]
with unitary operators $U(x)$ satisfying $U(0)=1$ and $U(x)U(y)=U(x+y)$.
The states of interest are the \bfi{regular states}, defined by 
\[
\<X,Y\>_x=\Tr(Y\rho(x) X^*)
\]
for some positive semidefinite Hermitian \bfi{density operator} 
$\rho(x)\in\ol\Qz(Z)$ with $\Tr\rho(x)=1$. In this case, the uncertain 
values
\[
\<X\>_x=\Tr(X\rho(x))
\]
viewed from $x\in\Rz^d$ are the \bfi{q-expectations}\footnote{
Traditionally, $\<X\>$ is called the expectation value of $X$, but such 
a statistical interpretation is not needed, and is not even possible
when $X$ has no spectral resolution.
} 
of $X$, and the uncertainty can be expressed of q-expectations, too.
For Hermitian $X$, it is given by
\[
\sigma_X(x)=\sqrt{\<X^2\>_x-\ol X(x)^2}.
\]

\subsection{The coherent variational principle}\label{ss.var}

A basic principle is the Dirac--Frenkel approach for reducing a 
nonrelativistic quantum problem to an associated classical 
approximation problem. In this approach, the variational principle for 
classical Lagrangian systems is rewritten for the present situation and 
then called the \bfi{Dirac--Frenkel variational principle}. It was first
used by \sca{Dirac} \cite{Dir.var} and \sca{Frenkel} \cite{Fre},
and found numerous applications; a geometric treatment is given in
\sca{Kramer \& Saraceno} \cite{KraS}. The action takes the form
\lbeq{eq.DiracFrenkel}
I(\psi) ~=~ \int\!dt~ \psi^* (i\hbar\partial_t - H) \psi
	~=~ \int\!dt~\Big(i\hbar\psi^*\dot\psi - \psi^*H\psi\Big)
\eeq
where the \bfi{quantum Hamiltonian} $H\in \Linx \Hz$ is a self-adjoint
operator. The coherent 1-form $\theta$ may be interpreted as the
Lagrangian 1-form corresponding to the Dirac--Frenkel action.
The Legendre transform of the \bfi{Lagrangian}
\[
L(\psi):=i\hbar\psi^*\dot\psi - \psi^*H\psi
\]
is the corresponding classical Hamiltonian
\[
\<H\> = \psi^*H\psi.
\]
The Dirac--Frenkel action is stationary iff $\psi$ satisfies the
\bfi{Schr\"odinger equation}
\[
i\hbar\dot\psi = H\psi,
\]
If one has a coherent space $Z$ and $\Hz=\Qz(Z)$ a quantum space of $Z$,
one can restrict $\psi$ to coherent states, and we get an action
\[
I(z) ~=~ \int\!dt~ \<z|\,(i\hbar\partial_t - H)\,|z\> 
\]
for the path $z(t)$. This coherent variational principle has first been 
proposed by \sca{Klauder} \cite{Kla.III}. The variational principle for 
the action $I(z)$ defines an approximate classical Lagrangian (and 
hence conservative) dynamics for the parameter vector $z(t)$. 
This \bfi{coherent dynamics} on $Z$ is regarded as a semiclassical 
(or \bfi{semiquantal}) approximation of the quantum dynamics. 
In two important cases, the norm of the state is preserved by the 
coherent dynamics -- $Z$ must either be \bfi{normalized}, i.e., 
$K(z,z)=1$ for all $z\in Z$, or projective (as defined in Subsection 
\ref{ss.geomQuant}).
The approximation turns out to be exact when the Hamiltonian belongs to 
the infinitesimal Lie algebra of the symmetry group of the coherent 
state. It is inexact but good if it is not too far from such an element.

The classical problem created by the Dirac--Frenkel approach is
again conservative, based an a classical action, which may or may not
be transformable into an equivalent Hamiltonian problem. The latter
depends on whether the Dirac--Frenkel Lagrangian is regular or singular.
Thus it is important that one understands the structure of classical 
singular Lagrangian problems.

\subsection{Coherent numerical quantum physics}

The Dirac--Frenkel variational principle is the basis of much of 
traditional numerical quantum mechanics, which heavily relies on 
variational methods. It plays an important role in approximation 
schemes for the dynamics of quantum systems.
In many cases, a viable approximation is obtained by restricting the
state vectors $\psi(t)$ to a linear or nonlinear manifold of
easily manageable states $|z\>$ parameterized by classical parameters 
$z$ which can often be given a physical meaning. 

What is commonly called a mean field theory is just the simplest
coherent state approximation. This is already much better than a
classical limit view, and in particular corrects for the missing zero
point energy terms in the latter.

An important application of this situation are the
\bfi{time-dependent Hartree--Fock equations} (see, e.g., 
\sca{McLachlan \& Ball} \cite{McLachB}), obtained by choosing $Z$ to 
be a Grassmann space.\footnote{
A Grassmann space is a manifold of all $k$-dimensional subspaces of a
vector space. It is one of the symmetric spaces.
} 
This gives the Hartree--Fock approximation, which is at the heart of
dynamical simulations in quantum chemistry. It can usually predict 
energy levels of molecules to within 5\% accuracy. Choosing $Z$ to be a 
larger space (obtained by the methods of Subsection \ref{ss.newOld})
enables one to achieve accuracies approaching 0.001\%.

Apart from Hartree-Fock calculations (symmetry group $U(N)$ on coadjoint
orbits of Slater states), this covers Hartree--Fock--Bogoliubov methods
(see, e.g., \sca{Goodman} \cite{Goo}), which include Bogoliubov 
transformations to get a quasiparticle picture
(symmetry group $SO(2N)$), and Gaussian methods (see, e.g., 
\sca{Pattanayak \& Schieve} \cite{PatS}, \sca{Ono \& Ando} \cite{OnoA})
used in quantum chemistry (symmetry group $ISp(2N)$). There are 
time-dependent versions of these, and extensions that go beyond the 
mean field picture, using either Hill-Wheeler equations in the 
generator coordinate method (see, e.g., \sca{Griffin \& Wheeler} 
\cite{GriW}) or coupled cluster expansions (see, e.g.,
\sca{Bartlett \& Musial} \cite{BarlM}) around the mean field. 

\subsection{Coherent chaos}

Since the Dirac--Frenkel variational principle gives a reduced 
deterministic dynamics for $z(t)$, it fits in naturally with the 
thermal interpretation. As discuseed in Subsection 4.2 of 
Part III \cite{Neu.IIIfound}, it is one of the ways to obtain a 
coarse-grained approximate dynamics for a set of relevant beable, in 
this case the $z\in Z$ labeling coherent states. 
Here we show that it can be used to study how -- in spite of the 
linearity of the Schr\"odinger equation -- chaos emerges through 
coarse-graining from the exact qauntum dynamics.

\sca{Zhang \& Feng} \cite{ZhaF} used the Dirac--Frenkel variational 
principle restricted to general coherent states to get a semiquantal
system of ordinary differential equations approximating the dynamics of 
the q-expectations of macroscopic operators of certain multiparticle 
systems. At high resolution, this deterministic dynamics is highly 
chaotic. This chaoticity is a general feature of approximation schemes 
for the dynamics of q-expectations or the associated reduced density 
functions. In particular, as discussed in detail in Part III 
\cite{Neu.IIIfound}, this seems enough to enforce the probabilistic 
nature of microscopic measurements using macroscopic devices. 

Zhang and Feng derive in a purely mathematical way -- without referring 
to probability or statistics -- the equations that they show to be 
chaotic. Thus what they do is completely independent of any particular 
interpretation of quantum physics.
They construct a semiclassical dynamics (where the relevant operators
are replaced by their q-expectations) and then discuss the resulting
system of ordinary differential equations. it turns out to be chaotic.
The exact quantum dynamics would be given instead by partial
differential equations!

In the overview of their paper, \sca{Zhang \& Feng} \cite[pp.4--9]{ZhaF}
state that they focus attention on understanding the question of 
quantum-classical correspondence (QCC), the search for an unambiguous 
classical limit, starting purely from quantum theory. They explore how,
under suitable conditions, classical chaos can emerge naturally from 
quantum theory. They use the semiquantal method discussed above for the 
exploration of the correspondence between quantum and classical dynamics
as well as quantum nonintegrability. They mentions the relations to 
geometric quantization and coherent states, and work in a group 
theoretic setting corresponding to coherent states defined by coadjoint 
orbits of semisimple Lie groups. 
Their coset space $G/H$ (or rather a complex line bundle over it arising
in geometric quantization and carrying some of the phase information)
discussed in \cite[pp.39]{ZhaF} is a coherent space with coherent 
product given by their (3.1.8). The variation of the effective quantum 
action in their (3.2.11) is the Dirac--Frenkel variational principle. 
The result of the variation is a symplectic system of differential
equations that has a semiclassical (or as they say, semiquantal) 
interpretation. This system gives an approximate dynamics for the 
q-expectations of the generators of the dynamical group. This dynamics 
is chaotic when the classical limit of the quantum system is not 
integrable.

\section{Field theory}\label{s.fieldTh}

This section defines the meaning of the notion of a field in the 
abstract setting of Section \ref{s.TILie} and shows how coherent spaces 
may be used to define relativistic quantum field theories. Nothing more
than basic definitions and properties are given; details will be given 
elsewhere.

\subsection{Fields}

In the geenral framework of Section \ref{s.TILie}, a \bfi{field} is an 
element $\phi$ of the space of $\Lz$-valued distributions 
$\Lz\otimes S(M,V)^*$ satisfying
\lbeq{e.dphi}
\frac{\partial}{\partial x_\nu} \phi(x)=p_\nu\lp\phi(x).
\eeq
Here $S(M,V)^*$ is the dual of the Schwartz space $S(M,V)$ of rapidly 
decaying smooth functions on $M$ with values in $V$ (or the space of 
$\Lz$-valued sections of a corresponding fiber bundle with generic 
fiber $V$).
Thus the \bfi{smeared fields} 
\[
\phi(f):=\int_M dx f(x)\phi(x),
\]
defined for arbitrary test functions $f\in S(M,V)$, provide quantities 
in $\Lz$. The primary beables are the distribution-valued q-expectations
\[
\phi_\cl(x)=\<\phi(x)\>_0
\]
of fields and the distribution-valued \bfi{Greens functions} 
\[
W(x,y)=\<\phi(x),\phi(y)\>_0
\]
of field products at some fixed spacetime origin $0$. After smearing 
with test functions, these distributions produce proper beables. 
A comparison of \gzit{e.dphi} with \gzit{e.momentum} shows that 
\[
\phi(x+z)=\phi(x)^{T_z}.
\]
As a consequence, field expectations from  different spacetime views 
satisfy
\[
\<\phi(x)\>_z=\<\phi(x+h)\>_{z-h},
\] 
showing that the choice of a fixed origin is inessential; a change of 
origin only amounts to a spacetime translation.

We also see that for any quantity $A\in\Lz$, the definition 
\[
A(x):=A^{T_x} \for x\in M
\]
defines a field. These fields are more regular than the fields occurring
in relativistic quantum field theory, which are proper distributions.

\subsection{Coherent spaces for quantum field theory}\label{ss.cohQFT}

The techniques of geometric quantization do not easily extend (except 
on a case by case basis) to the quantization of infinite-dimensional 
manifolds, which would be necessary for modeling quantum field theories.
However, the coherent space approach extends to quantum field theory.
The coherent manifolds are now infinite-dimensional, and their
topology is more technical to cope with than in the finite-dimensional
case. The process of second quantization is such an example of 
quantization of infinitely many degrees of freedom. Thus second 
quantized calculations become tractable via infinite-dimensional 
coherent spaces. For example the calculus of creation and annihilation 
operators was developed in \sca{Neumaier \& Ghaani Farashahi} 
\cite{NeuF.cohQuant} in terms of Klauder spaces, giving simple proofs 
of many standard results on calculations in Fock spaces.

The groups that can be most easily quantized are infinite-dimensional 
analogues of the symplectic, orthogonal and (for fixed particle number) 
unitary groups, Kac--Moody groups, some related groups, and their 
abelian extensions.
For example, the homogeneous quadratic expressions in finitely many
creation and annihilation operators form a symplectic Lie algebra
in the Boson case (CCR) and an orthogonal Lie algebra in the Fermion
case (CAR); see \sca{Zhang} et al. \cite{ZhaFG}.

This explains why
knowing the representation theory of these groups (in the form of
implications for their coherent spaces) is important.

Free quantum field theories are
essentially the large $N$ limit of the finite case. Large $N$ amounts
to discretizing configuration space or momentum space, keeping only $N$
degrees of freedom. This is the basis of lattice methods. The
thermodynamic limit $N\to \infty$ creates convergence problems -- one
has to struggle to avoid undefined expressions producing the infamous
''infinities''. The correct way to do this
requires some functional analysis and introduces cocycles (that, for
finite $N$, are trivial and hence can be avoided). For actual
calculations (by computer), one needs everything as explicitly as
possible, and coherent spaces yield explicit formulas for the things
of interest.

In quantum field theory one needs to take the limit
analytically rather than numerically, and a key problem is to decide
when these limits exists and whether one can find them explicitly
enough to get useful conclusions.
Using these formulas allows one to replace the usual long-winded
calculations with operators in the second-quantized formalism by fairly
short arguments.

Some work on infinite-dimensional versions is available;
in particular, for Fermions one needs the spin representation of
infinite-dimensional orthogonal groups, constructed in terms of
Pfaffians. The paper \sca{Gracia-Bond\'ia \& V\'arilly} \cite{GraV},
though not very readable, contains lots of details (but not in terms of
coherent spaces), and shows that the representation theory is enough to
settle the case of QED in an external field. This is easier than full
QED since the field equations are linear. The mathematical challenge is
the extension to nonlinear fields. (In \cite{GraV}, applications to the
nonlinear case are promised for a follow-up paper, but I could not find
any such paper.) 
In QED proper, asymptotic electrons are infraparticles rather than 
standard massive particles. Though we do not have a conventional Fock
 space, the asymptotic structure of QED is reasonably well understood. 
See, e.g., the work by \sca{Herdegen} \cite{Her1,Her2} and 
\sca{Kapec} et al. \cite{KapPRS}. 

\bigskip
{\bf Measures in infinite dimensions.}
For the quantization of infinite-dimensional manifolds, the Hilbert
space is traditionally constructed as a space of integrable functions
with respect to a measure on the manifold. Constructing the right
measure is difficult since there is no translation invariant
measure that could take the place of Lebesgue measure in finite
dimensions. Thus geometric quantization becomes an ad hoc procedure in
each particular case.
On the other hand, the coherent space approach generalizes
without severe problems to infinite dimensions.
Second quantization thus appears as the theory
of highest weight representations of infinite-dimensional Lie groups,
or rather its coherent space version, which is somewhat simpler to
manage. That it works in 2 dimensional spacetime is illustrated by the
success of conformal field theory which has a rigorous mathematical
description in terms of highest weight representations of the Virasoro
group.

This interpretation of the quantum world in terms of the classical is
important in quantum field theory when it comes to the explanation of
perturbatively inaccessible phenomena such as particle states
corresponding to solitons, or tunneling effects related to instantons.
See \sca{Jackiw} \cite{Jac}; however, his explanations are
mathematically vague. A coherent space setting makes this
mathematically rigorous, at least in the semiclassical approximation.

\bigskip
{\bf Form factors.}
Form factors appear as coefficients of operators in the algebra of
quadratics in the defining fields satisfying a conservation law
(i.e., vanishing divergence). They determine the possible interactions
with gauge fields.
A good description of form factors requires a detailed knowledge of the
causal irreducible representations of the Poincare group. See, e.g., 
{\sc Weinberg} \cite{Wei1,Wei2}, and \sca{Klink} \cite{Kli}; the results
there are not manifestly covariant. The coherent space approach
can be used to give nicer, manifestly covariant formulas. The form 
factors of a theory are needed for a subsequent analysis of spectral 
properties, such as the Lamb shift in QED.

\bigskip
{\bf Causal coherent manifolds.}
A \bfi{spacetime} is a smooth real manifold $M$ with a
Lie group $\Gz(M)$ of distinguished diffeomorphisms called
\bfi{spacetime symmetries} and a symmetric, irreflexive
\bfi{causality relation} $\times$ on $M$ preserved by $\Gz(M)$.
We say that two sections $j,k$ of a vector bundle over $M$ are
\bfi{causally independent} and write this as $j\times k$ if
\[
x \times y \for x\in \supp j,~y\in \supp k.
\]
Here $\supp j$ denotes the support of the function $j$.

A \bfi{causal coherent manifold} over a spacetime $M$ is a coherent
manifold $Z$ with the following properties:
\\
(i) The points of $Z$ form a vector space of smooth sections
of a vector bundle over $M$.
\\
(ii) The symmetries in $\Gz(M)$ act as unitary coherent maps.
\\
(iii) The coherent product satisfies the following causality conditions:
\lbeq{e.normal}
K(j,j')=1 ~~~\mbox{if } j \times j' \mbox{~or~} j~||~j'
\eeq
\lbeq{e.causal}
K(j+k,j'+k)=K(j,j') ~~~\mbox{if } j \times k \times j'.
\eeq

Examples of important spacetimes include:
\\
(i) \bfi{Minkowski spacetime} $M=\Rz^{1\times d}$ with a Lorentzian
inner product of signature $(+^1,-^d)$ and $x \times y$ iff $(x-y)^2<0$.
Here $d$ is the number of spatial dimensions; most often $d\in\{1,3\}$.
$\Gz(M)$ is the \bfi{Poincar\'e group} $ISO(1,d)$.
\\
(ii) \bfi{Euclidean spacetime} $M$ with $x \times y$ iff $x \ne y$.
Two Euclidean cases are of particular interest:
\\
(iii) For \bfi{Euclidean field theory}, $M=\Rz^4$ and $\Gz(M)$ is the
group $ISO(4)$ of Euclidean motions.
\\
(iv) For \bfi{chiral conformal field theory}, $M$ is the unit circle and
$\Gz(M)$ is the \bfi{Virasoro group}. Its center acts trivially on $M$
but not necessarily on bundles over $M$.

To give examples of a causal coherent manifold, we mention that from
any Hermitian quantum field $\phi$ of a relativistic quantum
field theory satisfying the \bfi{Wightman axioms}, for which the
smeared fields $\phi(j)$ (with suitable smooth real test functions $j$)
are self-adjoint operators, and any associated state $\<\cdot\>$,
the definition
\[
K(j,j'):=\<e^{-i\phi(j)}e^{i\phi(j')}\>
\]
defines a causal coherent manifold.

There are many known classes of relativistic quantum field theories
satisfying these properties in 2 and 3 spacetime dimensions. Under 
additional conditions one can conversely derive from a causal coherent 
manifold the Wightman axioms for an associated quantum field theory.
In 4 spacetime dimensions, only free and quasifree examples satisfying 
the Wightman axioms are known. The question of the existence of 
interacting relativistic quantum field theories in 4 spacetime 
dimensions is completely open.

Many tools from finite-dimensional analysis, in particular the Lebesgue
integral, Liouville measure, and averaging over compact sets must be
replaced by more unwieldy constructs, and limits need much more careful
considerations. Lack of heeding this would lead to the familiar
ultraviolet (UV) divergences and infrared (IR) divergences of
conventional quantum field theories. The IR and UV divergences go away
if the mathematically rigorous and correct considerations are applied.
This can be seen in quantum field theories in 2 and 3 spacetime
dimensions.

It is an open problem how to achieve the same in interacting quantum
field theories in the most important case, 4-dimensional spacetime.
There it is only known how to avoid the UV divergences, using careful
distribution splitting techniques in the context of causal perturbation 
theory. However, this approach only gives constructions for asymptotic 
series and misses the nonperturbative contributions needed for a fully 
defined interacting relativistic quantum field theories in 4 spacetime 
dimensions.

\bigskip

\end{document}